\documentclass[aps,pre,twocolumn,superscriptaddress]{revtex4-1}
\usepackage{amsfonts}
\usepackage{epsfig}
\usepackage{amsmath,amssymb}
\usepackage{times}
\usepackage{xcolor}
\usepackage{multirow}

\begin{document}

\title{Impacts of Opinion Leaders on Social Contagions}

\date{\today}
\author{Quan-Hui Liu}
\email{quanhuiliu8@gmail.com}
\affiliation{Web Sciences Center, School of Computer Science and Engineering, University of Electronic Science and Technology of China, Chengdu 611731, China}
\affiliation{Laboratory for the Modeling of Biological and Socio-technical Systems, Northeastern University, Boston, MA 02115, USA}
\affiliation{Big Data Research Center, University of Electronic Science and Technology of China, Chengdu 611731, China}

\author{Feng-Mao L\"u}
\affiliation{Web Sciences Center, School of Computer Science and Engineering, University of Electronic Science and Technology of China, Chengdu 611731, China}
\affiliation{Big Data Research Center, University of Electronic Science and Technology of China, Chengdu 611731, China}

\author{Qian Zhang}
\affiliation{Laboratory for the Modeling of Biological and Socio-technical Systems, Northeastern University, Boston, MA 02115, USA}

\author{Ming Tang}
\email{tangminghan007@gmail.com}
\affiliation{School of Information Science Technology, East China Normal University, Shanghai 200241, China}
\affiliation{Web Sciences Center, School of Computer Science and Engineering, University of Electronic Science and Technology of China, Chengdu 611731, China}
\affiliation{Big Data Research Center, University of Electronic Science and Technology of China, Chengdu 611731, China}

\author{Tao Zhou}
\affiliation{Web Sciences Center, School of Computer Science and Engineering, University of Electronic Science and Technology of China, Chengdu 611731, China}
\affiliation{Big Data Research Center, University of Electronic Science and Technology of China, Chengdu 611731, China}
\affiliation{Institute of Fundamental and Frontier Science, University of Electronic Science and Technology of China, Chengdu 611731, China}

\begin{abstract}

Opinion leaders are ubiquitous in both online and offline social networks, but the impacts of opinion leaders on social behavior contagions are still not fully understood, especially by using a mathematical model. Here we generalize the classical Watts threshold model and address the influences of the opinion leaders, where an individual adopts a new behavior if one of his/her opinion leaders adopts the behavior. First, we choose the opinion leaders randomly from all individuals in the network and find the impacts of opinion leaders make other individuals adopt the behavior more easily. Specifically, the existence of opinion leaders reduces the lowest mean degree of the network required for the global behavior adoption, and increases the highest mean degree of the network that the global behavior adoption can occur. Besides, the introduction of opinion leaders accelerates the behavior adoption, but does not change the adoption order of individuals. The developed theoretical predictions agree with the simulation results. Second, we randomly choose the opinion leaders from the top $h\%$ of the highest degree individuals, and find an optimal $h\%$ for the network with the lowest mean degree that the global behavior adoption can occur. Meanwhile, the influences of opinion leaders on accelerating the adoption of behaviors become less significant and can even be ignored when reducing the value of $h\%$.

\end{abstract}

\maketitle

\noindent\textbf{Studying social behavior spreading on complex networks is a relevant topic in the social science. Peer influences originated from adopted neighbors can induce the susceptible individuals to adopt the behavior, and for further may lead to the global behavior adoption if the condition meets. Recently, the influences of opinion leaders in social contagion have attracted many attentions, and the question about what impacts will the opinion leaders bring is raised. In this study, we propose a generalized Watts threshold model which addresses the influences of opinion leaders on adopting social behavior. A general result is that the introduction of opinion leaders allows other individuals in the network adopt the behavior more easily. Besides, when the opinion leaders are randomly chosen from the top $h\%$ of the highest degree individuals, i.e., the opinion leaders having high mean degree, we find an optimal $h\%$ for the network with the lowest mean degree that the global behavior adoption can occur. Our study also reveals some other interesting insights, such as the opinion leaders can accelerate the adoption of behaviors, but increasing the mean degree of opinion leaders weakens this impact.}

\section{Introduction} \label{sec:Introduction}

Social contagion processes exist widely in human society. Examples include the propagation of an opinion, the diffusion of an innovation and the adoption of a behavior. With the rapid development of technologies, these processes become an integral part of our everyday life and attract more and more attention from the researchers in the area of network science~\cite{Barrat:2008, Newman:2010,Castellano:2009, Pastor-Satorras:2015, WangWei:2017}.

To understand the underlying mechanisms of the real social spreading phenomenon, many mathematical models have been established. The well-known model used to describe the social contagion is the threshold model~\cite{Granovetter:1973, Centola:2007,Watts:2002}, where an individual adopts a new behavior if the number~\cite{Granovetter:1973, Centola:2007} (i.e., Granovetter model and Centola-Macy model ) or the fraction~\cite{Watts:2002} (i.e., Watts threshold model) of his/her adopted neighbors exceeds the adoption threshold. A remarkable phenomenon revealed by the Watts threshold model is that the final density of adopted individuals grows continuously and then decreases discontinuously as the mean degree of the network is increased~\cite{Watts:2002}. Meanwhile, the global behavior adoption occurs more easily in the network when the heterogeneity of the distribution for individuals' adoption thresholds is increased~\cite{Watts:2002}. Within the threshold model, the effects of network structure including the clustering coefficient~\cite{Hackett:2011, Whitney:2010}, the community structure~\cite{Gleeson:2008,Nematzadeh:2014}, weight~\cite{Hurd:2013}, multiplicity~\cite{Yagan:2012, Brummitt:2012, Lee:2014} and the temporal pattern~\cite{Takaguchi:2013} on the social contagion process have been investigated. Some other factors such as the initial seed size~\cite{Glesson:2007}, degree-dependent adoption thresholds~\cite{Lee:2017}, trend-driven~\cite{Kobayashi:2015}, spontaneous adoption~\cite{Ruan:2015} and persuasion~\cite{Huang:2016} also can affect the social contagion process. Besides, when the social reinforcement effect~\cite{Lu:2011, Zheng:2013, Liu:2016,Liu:2017} originated from the cumulative exposure of the behaviors is considered, a three-stated non-Markovian social contagion model is also built in Ref.~\cite{Wang:2015}. 

A common assumption in the Watts threshold model is that the probability for an individual to adopt a new behavior is only determined by the state of his/her direct neighbors. In real-world, the factors which impact individuals on adopting a new behavior or enrolling an activity are not only from the peer influences, but also from a wide variety of sources~\cite{Zhang:2013, Hodas:2014, Cozzo:2013}, such as the synergy of having adopted other behaviors~\cite{Liu:2018-1, Liu:2018-2}, the mass media~\cite{Walter:2014,Kobayashi:2015}, etc. Specifically, there is a kind of individuals defined as the opinion leaders, who have great influences in the opinions, attitudes, motivations and behaviors of others~\cite{Rogers:1962,Myers:1972,Richins:1988,Iyengar:2011, Lu:2016}. Since the opinion leaders exist widely in real world and they play a great role when a person is in making decision. There are many studies on the impact of opinion leaders in social contagion, ranging from the experiments to the data analysis and the modeling. In the early time, many experiments have shown that the opinion leaders are very effective in decreasing the rate of unsafe sexual practices~\cite{kelly:1991}, cesarean births~\cite{lomas:1991} and promotion of mammography screening~\cite{Earp:2002}. With the development of Internet, many datasets are available for studying the online human behaviors. Such as the authors in Ref.~\cite{Park:2013} find that the opinion leaders on Twitter play a significant role in arousing individuals who have social surveillance motivations and in leading them to actively use Twitter. Besides, the opinion leaders encourage their followers to enroll an election, to participate a protest activity and other public political process by using the Twitter~\cite{Park:2013}. In the aspect of modeling evolutional game modeling, a repeated public goods game is employed and the authors find that the effort of leaders can reduce the likelihood that cooperation fails~\cite{Hooper:2010}. For the process of consensus-based decision-making, the groups benefit a lot from the leaders when there are the time pressures and significant conflicts of interest between members in the group~\cite{Gavrilets:2016}. In modeling the diffusion of technological innovation~\cite{Rogers:2010}, the threshold model is adopted in Ref.~\cite{Cho:2012} and the authors study the impacts of different ways to choose the opinion leaders. In their model, the opinion leaders are set as the initial adopted seeds.They find that the opinion leaders with high sociality are the best choosing method for fast diffusion, and those with high distance centrality are the best way for the maximum cumulative number of adopters~\cite{Cho:2012}. In studying the opinion leaders' role in the adoption of new products, the authors in Ref.~\cite{Eck:2011} propose a model which considers individuals' preference and social influence. Specially, the social influence quantified by the fraction of adopted neighbors is not included in their model directly. Instead, it is regarded as a continuum, which the 
opinion leader with a small weight on this continuum compared with the generally individuals. They find the effects of  opinion leaders increase the speed of contagion and the maximum adoption fraction~\cite{Eck:2011}. Though many studies have been investigated on the impacts of opinion leaders in social contagion, a mathematic model which focuses on the impacts of the single individual's opinion leaders on his/her behavior adoption is still non-existent. 

In this paper, we articulate a generalized threshold model to explore the impacts of opinion leaders on social behavior contagion. In this model, we assume the distribution for the number of opinion leaders that an individual follows in the network is $Q(k_0)$, i.e., the probability for a random individual having $k_0$ opinion leaders. The network that we adopt to perform the contagion process is undirected and is produced by the uncorrelated configuration model~\cite{Catanzaro:2005}. A directed link is added between each follower and each of his/her leaders. The directed link is adopted since we may follow the celebrity, i.e., a kind of leader~\cite{Valente:2007}, in the online social networks (e.g., \emph{Twitter}, \emph{Weibo}), who will not follow us. But once a celebrity who we follow adopts a behavior, we will adopt it with very large probability. For the sake of simplicity, the mechanism of opinion leaders is introduced that once an opinion leader of an individual adopts the behavior, this individual adopts the behavior too, ignoring whether the fraction of his/her adopted neighbors exceeds the adoption threshold or not.

At first, for each individual and based on $Q(k_0)$, we randomly choose all other individuals from the network and assign them as the opinion leaders to this individual. In this scenario, the mean degree of the opinion leaders approximates the mean degree of the network. We find that the introduction of opinion leaders makes the behavior be adopted more easily. Specifically, the existence of opinion leaders reduces the lowest mean degree of the network required for the global behavior adoption, and also increases the highest mean degree of the network that the global behavior adoption can occur. Besides, the impacts of opinion leaders accelerate the adoption of behaviors, but do not change the adoption order of the individuals, which means the low- and mean-degree individuals are still responsible for triggering the global behavior adoption~\cite{Watts:2002}. The theoretical predictions match with the simulation results. Secondly, we consider the situation that the opinion leaders are well connected in real world, such as the celebrity in the \emph{Twitter} networks always having many followers. We randomly choose the individuals from top $h\%$ of the highest degree individuals and assign them as the opinion leaders to each individual based on $Q(k_0)$. Note that the opinion leaders will have higher mean degree when the value of $h\%$ is set to be smaller. We find an optimal  $h\%$ for the network with the lowest mean degree that the global behavior adoption can occur. For the networks with high mean degrees, decreasing the mean degree of opinion leaders (i.e., increasing $h\%$) can increase the highest mean degree of the network that the global behavior adoption can occur. The impacts of opinion leaders become less significant and even can be ignored in accelerating the behavior adoption when the mean degree of the opinion leaders is increased (i.e., $h\%$ is reduced)

The paper is organized as follows: Sec. II introduces the model. In Sec. III we develop the theory when the opinion leaders are randomly chosen from the network. The simulation results and theoretical predictions are presented in Sec. IV. In Sec. V we summarize the conclusions.\\

\section{Model} \label{sec:model}
We generalize the Watts threshold model and study the opinion leaders' impacts on the behavior adoption of their followers on complex networks~\cite{Newman:2010}. The uncorrelated configuration model~\cite{Catanzaro:2005} with a given degree distribution $p(k)$ is adopted to produce the network, where the degree-degree correlations can be neglected for large and sparse networks. Nodes in the network represent the individuals and links denote the interactions between individuals. The distribution for the number of opinion leaders that an individual follows is $Q(k_0)$. For the simplicity, we set the fraction of individuals with 3 or more than 3 opinion leaders as zero. And $Q_{0,1,2}(x, y, z)$ is used to denote the fractions of individuals with 0 opinion leader, 1 opinion leader and 2 opinion leaders respective as $x$, $y$ and $z$. Therein, $x\geq0, y\geq0, z\geq0$ and $x+y+z=1$. We consider two scenarios to choose the opinion leaders for each individual on the network. One is that we choose the individuals randomly and assign them as the opinion leaders to each individual. The second scenario is that the individuals who are going to be assigned as the leaders are chosen from the top $h\%$ of highest degree individuals. If two or more individuals are with the same degree, they are randomly ranked. Note that a directed link is added between each individual and each opinion leader of him/her. With this directed link, an individual can know whether his/her opinion leaders adopt the behavior or not. In the computation of the mean degree of the network and the fraction of adopted neighbors for an individual, only the undirected links are considered.

Initially, $\rho_0$ fraction of individuals are randomly chosen as the adopted individuals and the remaining individuals are set in the susceptible state.  During the adoption process of the behavior, all individuals will be in one of the following two states, the susceptible state (S) and adopted state (A). The dynamical process evolves as each individual with degree $k$ changes its state from S to A if one of his/her opinion leader adopts the behavior or the fraction of its adopted neighbors exceeds the adoption threshold. The dynamics terminates when there is no adoption process.

\section{Theory For Random Opinion Leaders} \label{sec:theory}
We use the treelike approximation method~\cite{Glesson:2007} to derive the final adoption density denoted by $\rho_{\infty}$ when the opinion leaders are randomly chosen from all individuals in the network. As described in the model, there are two cases that an individual will adopt the behavior. The first is that if one of his/her opinion leaders adopts the behavior. The second is that the fraction of adopted neighbors exceeds the adoption threshold. Let $\rho_t$ represent the fraction of adopted individuals at time $t$, and $q_t$ be the probability that a random neighbor of an individual is in the adopted state at time $t$. Thus, for an individual with susceptible state initially, the probability that he/she adopts the behavior at time $t$ only caused by his/her opinion leaders is calculated as
\begin{eqnarray}\label{lt}
L_t=\sum_{k_0}Q(k_0)[1-{(1-\rho_{t-1})}^{k_0}].
\end{eqnarray}
Therein, ${(1-\rho_{t-1})}^{k_0}$ represents the probability that none of his/her $k_0$ opinion leaders adopt the behavior. $L_t$ is only correlated with $Q(k_0)$ and $\rho_{t-1}$.

For an individual of degree $k$ and only with the peer influences, the probability for him/her adopting the behavior at time $t$ is $\sum_{m=0}^{k}{k \choose m}{(q_{t-1})}^m{(1-q_{t-1})}^{(k-m)}F\big(\frac{m}{k}\big)$, therein, $F(x)$ denotes the probability that the adoption threshold of an individual is less than $x$. For the simplicity, we assume all individuals are with the same adoption threshold $T_0$. That is,
\begin{equation}\label{F}
F(x)=\left\{
\begin{array}{rcl}
{1} & {\mathrm{ x \geq T_0},}\\
{0} & {\mathrm {x <T_0}.}
\end{array} \right.
\end{equation}

Combining both the influences from opinion leaders and peers, the fraction of adopted individuals at time $t$ is computed as
\begin{eqnarray}\label{rhot}
\nonumber
\rho_t&=&\rho_0+(1-\rho_0)\bigg[L_{t}+(1-L_{t})\sum_k{p(k)}\\
&\times&\sum_{m=0}^{k}B_{k,m}(q_{t-1})F\bigg(\frac{m}{k}\bigg)\bigg],
\end{eqnarray}
where $B_{k,m}(q)$ is a binomial expression equal to ${ k \choose m}{(q)}^m{(1-q)}^{(k-m)}$. Using the similar derivation method, the probability that a random neighbor of a susceptible individual is in adopted state at time $t$ is
\begin{eqnarray}\label{qt}
\nonumber
q_t&=&\rho_0+(1-\rho_0)\bigg[L_{t}+(1-L_{t})\sum_k{\frac{k}{z}p(k)}\\
&\times&\sum_{m=0}^{k-1}B_{k-1,m}(q_{t-1})F\bigg(\frac{m}{k}\bigg)\bigg],
\end{eqnarray}
where $z$ is the average degree of the  network equal to $\sum_k{kp(k)}$. By iterating Eqs.~(\ref{rhot})
and (\ref{qt}) with $q_0=\rho_0$, one can determine $\rho_t$ for any $t>0$, and both $\rho_t$ and $q_t$ converges to $\rho_{\infty}$ and $q_{\infty}$, respectively, when $t\rightarrow\infty$. Thus, the final adoption density $\rho_{\infty}$ is determined.

Another interesting point is the critical condition that determines whether a global behavior adoption may occur or not. By linearizing Eq.~(\ref{qt}) near $q=0$, we can get the following equation (The detailed derivation is presented in the Appendix part.)
\begin{eqnarray}
\nonumber
q_t&\approx&\rho_0+(1-\rho_0)q_{t-1}\\
&\times&\bigg[\sum_{k_0}Q(k_0)k_0+\sum_k\frac{k(k-1)}{z}p(k)F\bigg(\frac{1}{k}\bigg)\bigg].
\end{eqnarray}

{\noindent}The condition that the global behavior adoption occurs is $q_t{\geq}q_{t-1}$ since this guarantees that $q_t$ increases with t. For infinite initial fraction of adopted individuals, we can approximate this critical condition as
\begin{eqnarray}\label{threshold}
\sum_{k_0}{k_0}Q(k_0)+\sum_k\frac{k(k-1)}{z}p(k)F(\frac{1}{k})\geq 1.
\end{eqnarray}
We will find when $Q(0)=1$, Eq.~(\ref{threshold}) reduces to the condition derived by Watts using percolation method~\cite{Watts:2002}.

\begin{figure}
\centering
\includegraphics[width=3in]{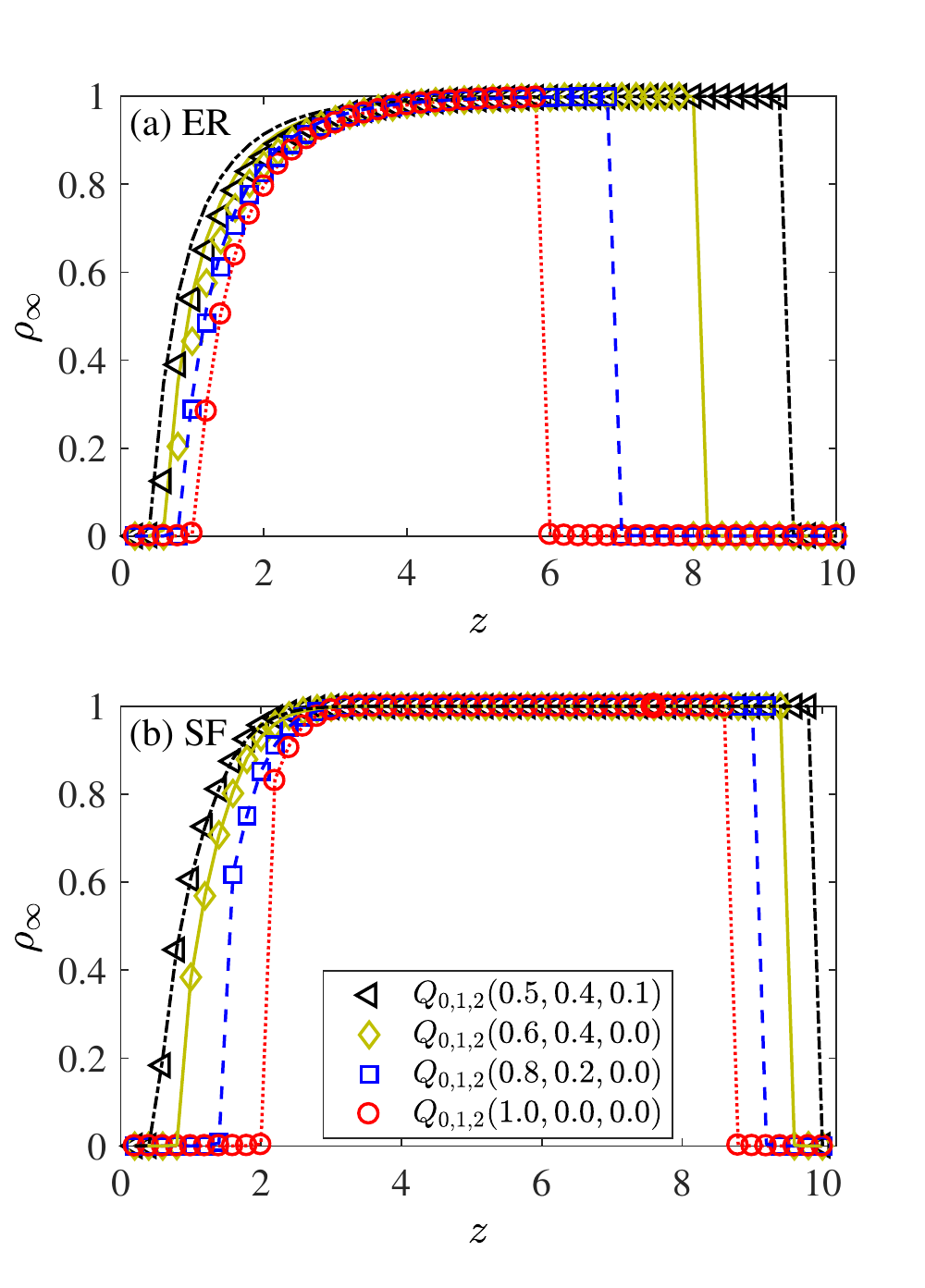}
\caption{(Color online) Final adoption density $\rho_{\infty}$ versus mean degree $z$ of ER (a) and SF (b) networks when opinion leaders are randomly chosen from all individuals in the network. Therein, $Q_{0,1,2}(0.5,0.4,0.1)$ represents the fractions of individuals in the network with $zero$ leader, $one$ leader and $two$ leaders leaders are respectively as 0.5, 0.4 and 0.1. It is the same for the meaning of $Q_{0,1,2}(0.6,0.4,0.0)$, $Q_{0,1,2}(0.8,0.2,0.0)$ and $Q_{0,1,2}(1.0,0.0,0.0)$. The symbols are the simulation results and the lines with the same color are the corresponding theoretical predictions from Eqs.~(\ref{rhot}) and~(\ref{qt}).}
\label{simThe}
\end{figure}

\section{NUMERICAL VERIFICATION}
In this section, we present the simulation results and the theoretical predictions. We perform the simulations on Erd\"{o}-E\'{e}nyi (ER)~\cite{ERDdS:1959} and scale-free (SF)~\cite{Catanzaro:2005} networks. For the SF network, the degree distribution is $P(k)={\Gamma}k^{-\gamma}$, where $\gamma$ is the degree exponent and the
coefficient is $\Gamma=1/\sum_{k_{min}}^{k_{max}}k^{-\gamma}$ with the
minimum degree $k_{min}=3$, maximum degree $k_{max}{\sim}N^{1/(\gamma-1)}$ and $\gamma=3.0$.  Unless otherwise specified, the network size is set as $N=10^5$ and the adoption threshold for all individuals are set as $T_0=0.18$. Initially, $\rho_0=10^{-4}$ fraction of individuals are randomly chosen and are set in the adopted state. The remaining individuals are in the susceptible state. At least $10^3$ independent dynamical realizations on a fixed network are used to calculate the pertinent average values, which are further averaged over 20 network realizations.

\begin{figure*}
\centering
\includegraphics[width=5in]{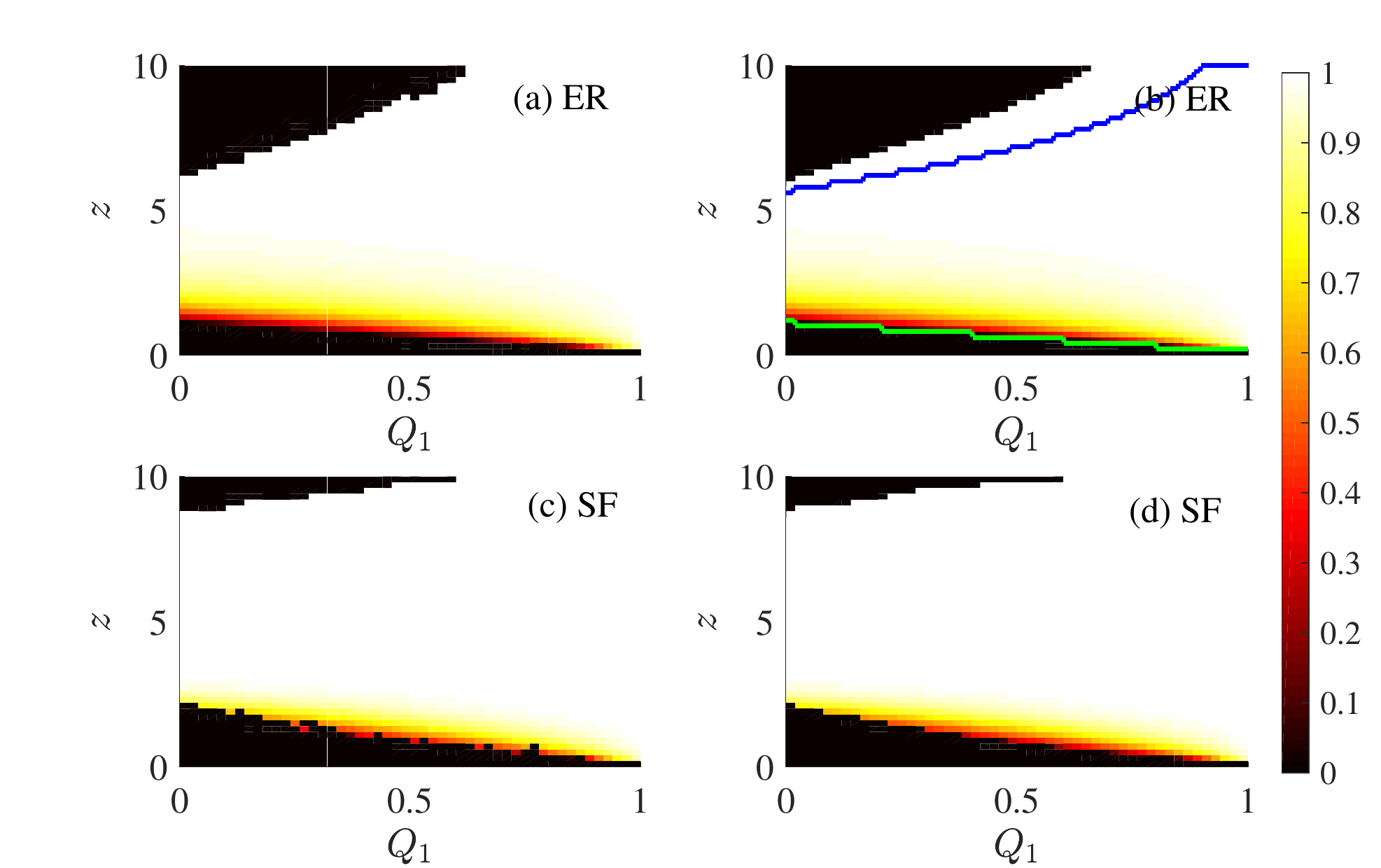}
\caption{(Color online)  Effect of the fraction of individuals with $one$ opinion leader ($Q_1$) and the mean degree of network ($z$) on the final adoption density for the opinion leaders randomly chosen in the network. (a) simulation results and (b) the theoretical predictions from Eqs.~(\ref{rhot}) and~(\ref{qt}) for the social contagion process on ER networks. (c) simulation results and (d) the theoretical predictions for the social contagion process on SF networks. The color-coded value represents the final adoption density. The cascade boundaries on subfigure (b) are from Eq.~(\ref{threshold}). The fraction of individuals with $zero$ leader is set as $Q_0=1-Q_1$ and the fraction of individuals with two and more than opinion leaders is set to $zero$. }
\label{global3D}
\end{figure*}

\subsection{Opinion Leaders with mean degree equal to network}

In this part, we present the simulation results and theoretical predictions for the scenario that the opinion leaders are randomly chosen in the network.

Figs.~\ref{simThe} (a) and (b) respectively present the final adoption densities versus the mean degree $z$ of ER and SF networks. From both kinds of networks, we find the impacts of opinion leaders make the behavior be adopted more easily. Specifically, the existence of opinion leaders not only reduces the lowest mean degree of the network required for the global behavior adoption, it also increases the highest mean degree of the network that the global cascade can occur. For example, in Fig.\ref{simThe} (a), when all individuals in the network have no opinion leaders, i.e., $Q_{0,1,2}(1,0,0)$, the model reduces to the classical Watts threshold model. The lowest mean degree of the ER network required for the global behavior adoption occurring is $z=1.0$. When the influences of opinion leaders are incorporated, e.g., $Q_{0,1,2}(0.8,0.2,0.0)$, it is equal to add the edges (i.e., the directed dependent edges between an individual and his/her opinion leaders) on the network which increases the practical mean degree of the network. Thus, the global adoption of behavior can occur even $z<1$. When the mean degree of the network is increased, it becomes difficult for individuals to adopt the behavior because they need more adopted neighbors to exceed the adoption threshold. Therefore, the final adoption density is suddenly decreased to zero~\cite{Watts:2002} when the mean degree of the network crosses a critical value (i.e., the highest mean degree). Additionally, with the influences of opinion leaders, increasing the fraction of individuals with the opinion leaders will enable the global behavior adoption occurring on the network with higher mean degree. These conclusions hold in SF networks as well [see Fig.\ref{simThe} (b) ]. The simulation results match well with the theoretical predictions.

Next, we investigate the impacts of opinion leaders on the final adoption density in the plane $(Q_1, z)$. $Q_1$ denotes the fraction of individuals in the network with $one$ opinion leader and the fraction of individuals with two and more than two opinion leaders is fixed as $zero$. Then the fraction of individuals with no opinion leaders is $1-Q_1$. We adopt the uniform adoption threshold for all individuals and set $T_0=0.18$. The color-coded values in Fig. 2 represent the final adoption densities.  Figs.~\ref{global3D} (a) and (b) respectively represent the simulation results and the theoretical predictions from Eqs.~(\ref{rhot}) and~(\ref{qt}) on ER networks. As we found in Fig.~\ref{simThe} that increasing the fraction of individuals with opinion leaders decreases the lowest mean degree required for the network that the global behavior adoption can occur, it also enlarges the highest mean degree of the network that the global cascade can persist. Here, we also find the same results as shown in Fig. 2. Meanwhile, the theoretical predictions agree well with simulation results. Again, similar phenomena can be observed in SF networks. Two lines in Fig.~\ref{global3D} (b) are the boundaries from Eq.~(\ref{threshold}). Although the line does not match well with the boundary for network with high mean degree, it can qualitatively reflect the trend of the highest mean degree of the network that the global cascade can persist when increasing $Q_1$. We didn't show the boundaries for the SF networks in Fig.~\ref{global3D} (d) since the degree heterogeneity causes the large fluctuation of the numerical results~\cite{Huang:2016}.

\begin{figure*}
\centering
\includegraphics[width=6in]{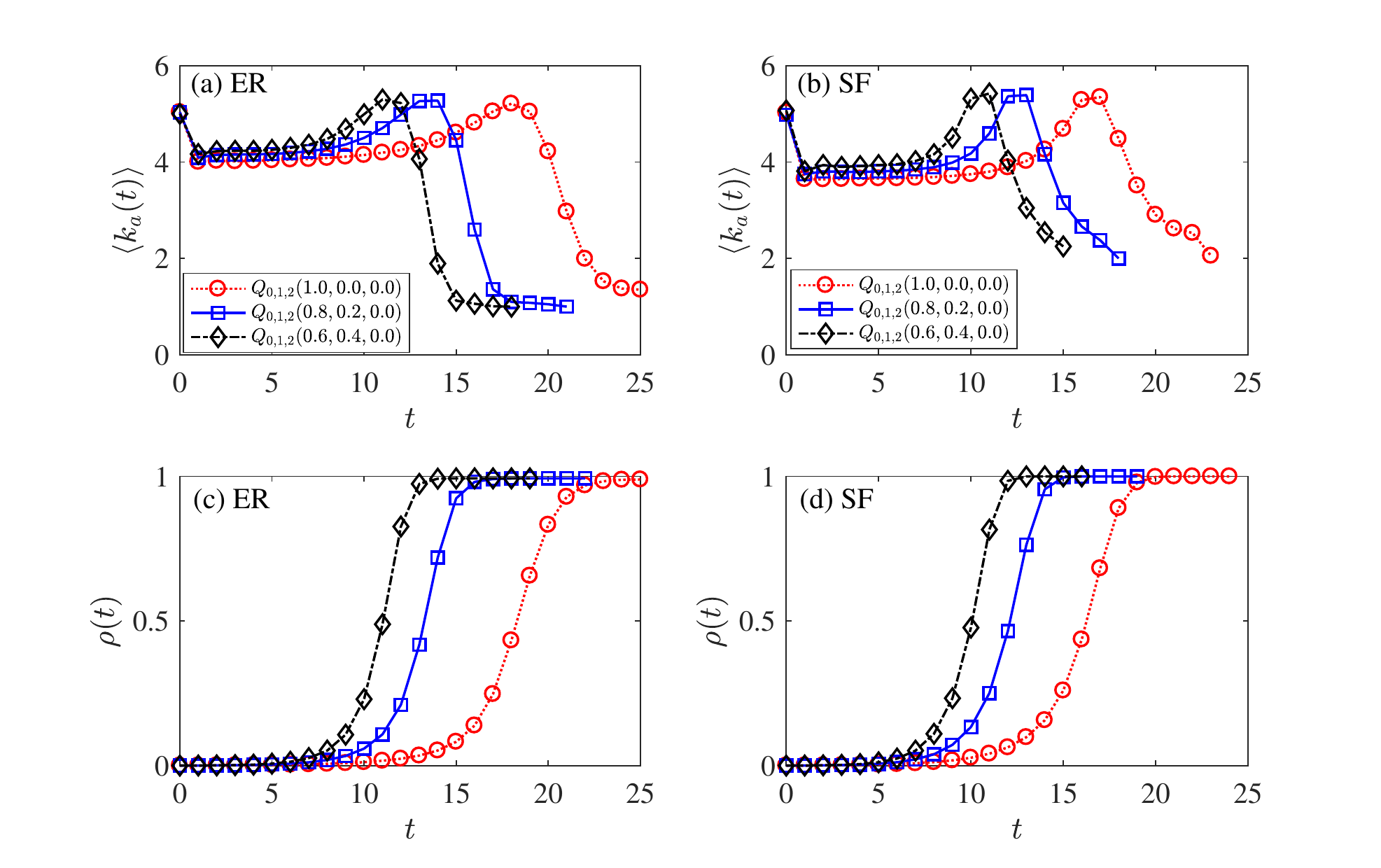}
\caption{(Color online)  Temporal plots of the mean degree of newly adopted individuals and the cumulative density of adopted nodes when the opinion leaders are randomly chosen in the network. The mean degree  $\langle k_a(t) \rangle$ of newly adopted individuals for ER (a)  and SF (b) networks. The cumulative density of adopted nodes on ER (c) and SF (d) networks. The parameters for the simulation in (c) [(d)] are the same as used in (a) [(b)]. Both the mean degree for ER and SF networks are set as $z=5.0$.}
\label{globalTime}
\end{figure*}

To get a deep understanding on how the opinion leaders of individuals impact the adoption process, we calculate the mean degree of newly adopted individuals and the cumulative density of adopted individuals versus time $t$, as shown in Fig.~\ref{globalTime}. For the behavior spreading with opinion leader [e.g., $Q_{0,1,2}(0.8,0.2,0.0)$] and without opinion leader [e.g., $Q_{0,1,2}(1.0,0.0,0.0)$], both them display the same trend, 
where low- and mean-degree individuals adopt the behavior first and then trigger the global behavior adoption. It implies that the opinion leaders will not change the adoption order of the individuals~\cite{Watts:2002}. The average degree of newly adopted individuals increases first and then decreases, which can be explained as follows. Initially, there are few adopted individuals. Compared with the individuals of high degree, the individuals with low degree are more likely to exceed the adoption threshold and then adopt the behavior first. With continuous behavior adoption, there are more and more adopted neighbors around the high degree individuals, which will lead them to adopt the behavior. Finally, the nodes in the periphery of the network are going to adopt the behavior.\\

Besides, in contrast with the Watts threshold model, i.e., $Q_{0,1,2}(1.0,0.0,0.0)$, the existence of opinion leaders can accelerate the spread of behaviors. It can be explained as follows. The behavior adoption of an opinion leader will lead her/his followers to adopt the behavior ignoring whether their fractions of adopted neighbors exceed the adoption thresholds or not, as shown in Figs.~{\ref{globalTime}} (c) and (d)]. Specifically, when comparing $Q_{0,1,2}(0.6,0.4,0.0)$ with $Q_{0,1,2}(0.8,0.2,0.0)$, we will find that the larger the fraction of individuals with the opinion leaders, the less time is needed for the behavior adoption approaches the final state. These conclusions hold for both ER and SF networks.

\begin{figure}
\centering
\includegraphics[width=3in]{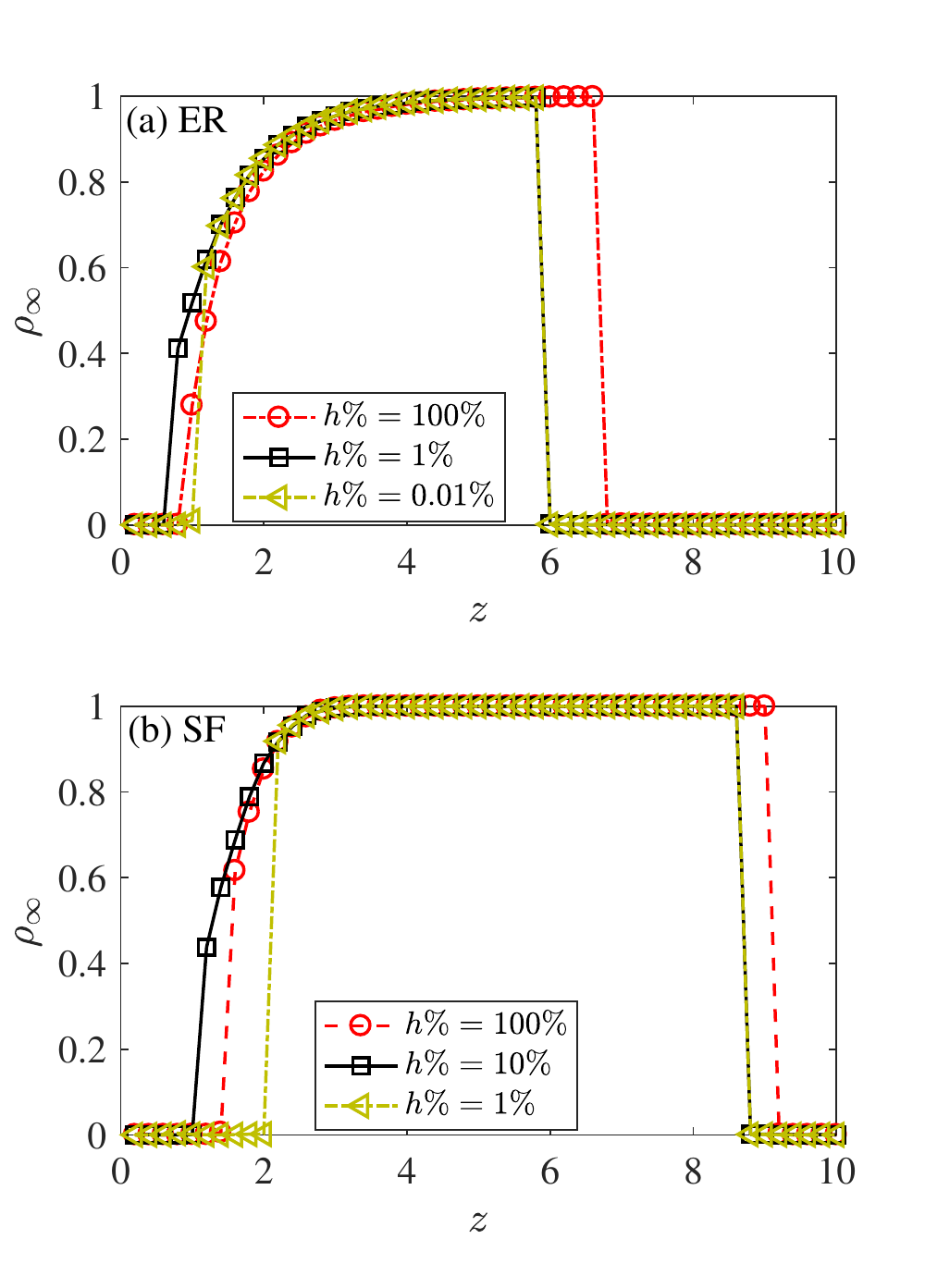}
\caption{(Color online)  Final adoption density $\rho_{\infty}$ of adopted nodes in ER (a) and SF (b) networks versus mean degree $z$ for different $h\%$. The distribution of number of leaders is set as $Q_{0,1,2}(0.8,0.2,0.0)$.}
\label{hubleader}
\end{figure}

\begin{figure}
\centering
\includegraphics[width=3in]{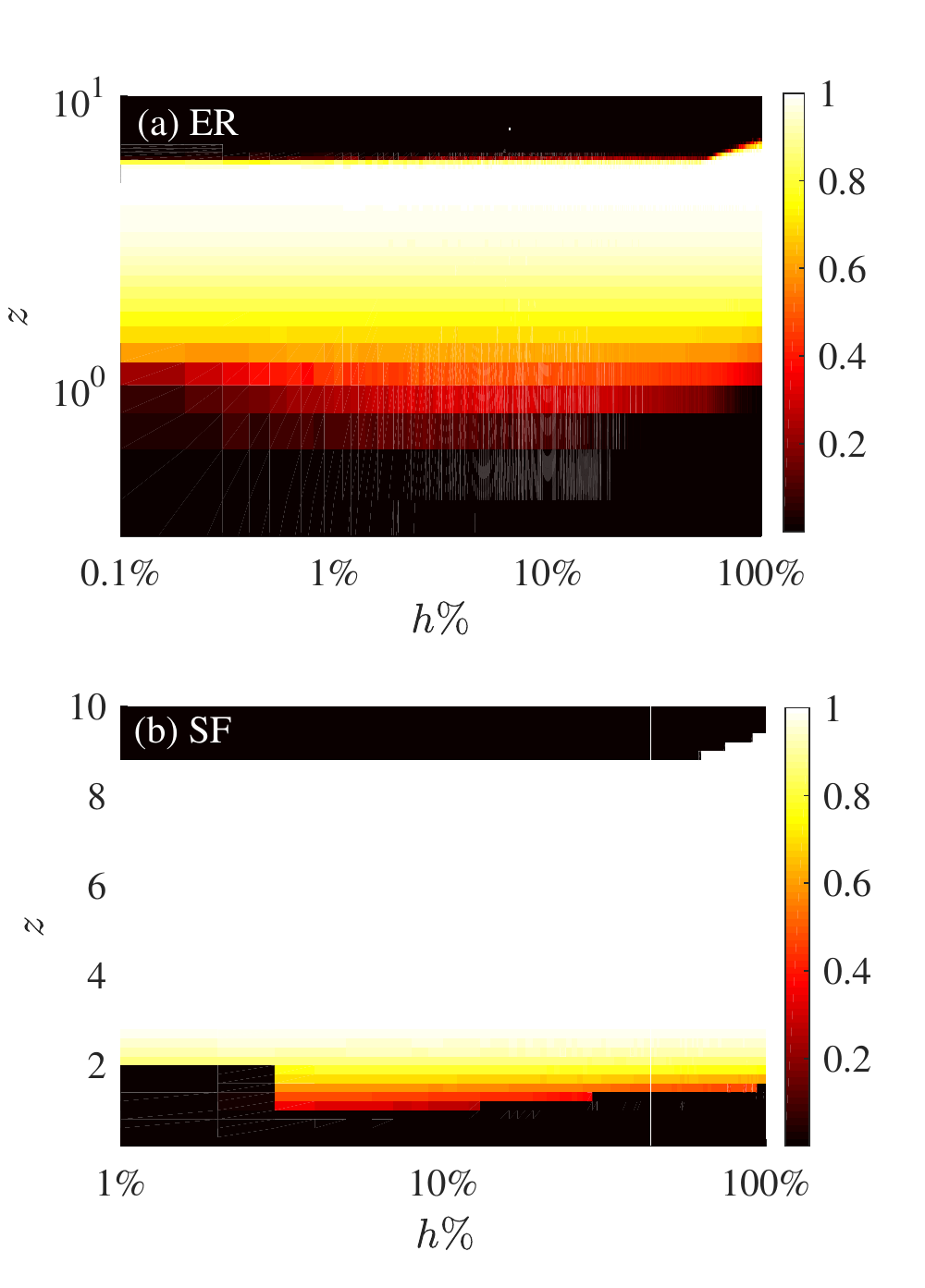}
\caption{(Color online) Effect of $h\%$ and the mean degree of ER network (a) and (b) SF network on the final adoption density. The distribution of number of leaders is set as $Q_{0,1,2}(0.8,0.2,0.0)$.}
\label{hub3D}
\end{figure}

\subsection{Opinion Leaders with high mean degree}

In the previous part, we chose the opinion leaders randomly from all individuals in the network. In such scenario, the mean degree of the opinion leaders is approximate to the mean degree of the network. In real world, the opinion leaders, such as the celebrity, are well connected. To consider this situation, in this part we investigate the impacts of opinion leaders on the adoption of behaviors when the opinion leaders are with high mean degree. Firstly, we rank the individuals by their degrees from high to low. If two or more individuals are with the same degree, they are randomly ranked. The individuals who are going to be assigned as the opinion leaders are randomly chosen from the top $h\%$ of the highest degree individuals. If $h\%=100\%$, the leaders are randomly chosen from all individuals which reduces to the previous part (i.e., the opinion leaders are with the mean degree equal to the mean degree of the network). The smaller value of $h\%$, the higher mean degree of the opinion leaders.

Firstly, we study the effect of mean degree of the network on the final adoption densities for different $h\%$. The fraction of individuals with $one$ opinion leader is fixed as $0.2$ and other individuals are with no opinion leaders. For the ER network with the low mean degree, as shown in Fig.~\ref{hubleader} (a), when the mean degree of the opinion leader is increased (i.e., decreasing $h\%$=100\% to $h\%$=0.01\%), we find the lowest mean degree of the network required for the global behavior adoption decreases first and then increases. It means there exists an optimal mean degree of the opinion leaders for the network on which the global behavior adoption can occur with the lowest mean degree. It can be explained as follows. If we randomly choose the opinion leaders from all individuals in the network, i.e., $h\%=100\%$, most of them will have degree equal to $one$. It's difficult for these opinion leaders to adopt the behavior since they are not in the giant component [Here the new added directed links between a follower and his/her opinion leaders are included in making up the giant componet].  Thus, the effects on stimulating their followers to adopt the behavior can be ignored. If we appropriately increase the mean degree of opinion leaders, i.e., $h\%=1\%$, the opinion leaders are more likely to be chosen from the giant component and are not with higher degrees (compared with $h\%=0.01\%$). In this case, the opinion leaders will adopt the behavior and also stimulate their followers to adopt the behavior. However, if the opinion leaders are with the higher mean degree, i.e., $h\%=0.01\%$. Though they are in the giant component, it is difficult to adopt the behavior because they need more adopted neighbors compared with $h\%=1\%$ to exceed the adoption threshold.  Meanwhile, as presented in Fig. 1, when the mean degree of the network is increased crossing a critical value, the final adoption density is suddenly decreased to \emph{zero}~\cite{Watts:2002}. Additionally, increasing the mean degree of the opinion leaders makes it difficult for the opinion leader themselves to adopt the behavior first. Thus, the highest mean degree of the network that the global behavior adoption can occur is decreased when the mean degree of the opinion leaders is increased. These results hold the same for the SF networks, as shown in Fig.~\ref{hubleader} (b).

We investigate the final adoption density in the plane $(h\%, z)$. As we reported for the simulation results in Fig.~\ref{hubleader}, decreasing $h\%$ from $100\%$ to a smaller value (i.e., $0.1\%$ for ER network and $1\%$ for SF network), the lowest mean degree required for the global adoption  decreases first and then increases, as shown in Fig.~\ref{hub3D}. For the network with high mean degree, increasing the value of $h\%$ enlargers the highest mean degree of the network that the global behavior adoption can occur. These results hold for both the ER and SF networks.

\begin{figure*}
\centering
\includegraphics[width=6in]{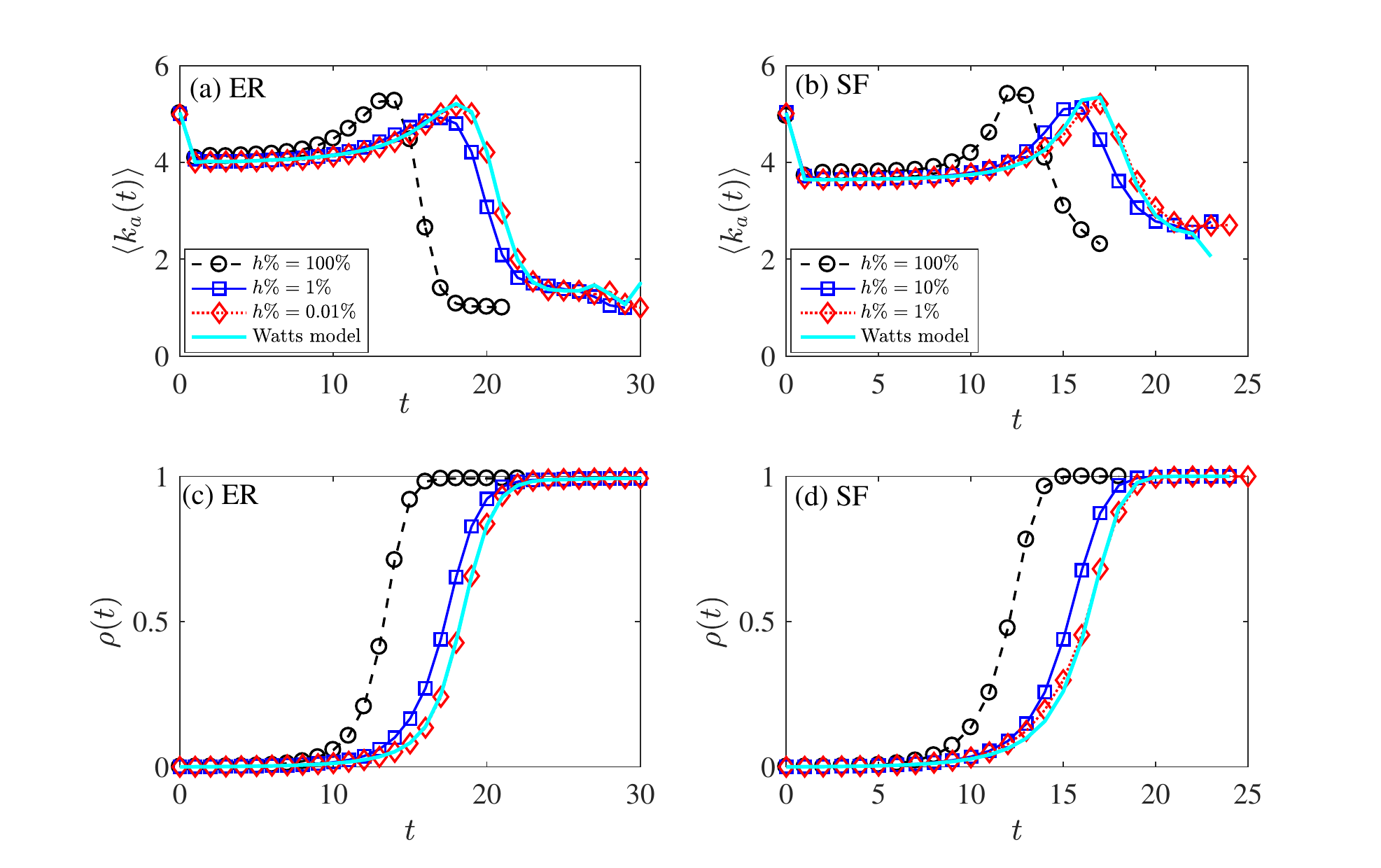}
\caption{(Color online) Temporal plots of the mean degree of newly adopted individuals and the cumulative density of adopted nodes for different $h\%$. The mean degree  $\langle k_a(t) \rangle$ of newly adopted individuals for ER (a)  and SF (b) networks. The cumulative density of adopted nodes on ER (c) and SF (d) networks. The parameters for the simulation in (c) [(d)] are the same as used in (a) [(b)]. The cyan line in each subfigure is the simulation results of the Watts threshold model with the same parameters. Both the mean degree for ER and SF networks are set as $z=5.0$.}
\label{hubTime}
\end{figure*}

Finally, we also calculate the mean degree of newly adopted individuals and the adoption density versus time $t$ in ER and SF networks, as shown in Fig.~\ref{hubTime}. First, we find that the opinion leaders of high degrees have no impact on the adoption order of individuals, which means the low- and mean- degrees individuals are still responsible for triggering the global behavior adoption. Besides, the higher the mean degree of the opinion leaders, the less the influences of opinion leaders in accelerating the adoption of behavior, since it becomes difficult for the opinion leaders to have enough adopted neighbors to exceed the adoption threshold. Specifically, when $h\%=0.01\%$, the impacts of opinion leaders even can be ignored, e.g., the line for $h\%=0.01\%$ overlapping with the line of Watts threshold model in Figs. 6 (a) and (c).

\section{Conclusion}
Social behavior spreading is a kind of complex contagion since there are many factors nonlinearly impacting the adoption of behavior. In this paper, we purpose a generalized Watts threshold model incorporating the influences of opinion leaders in the behavior adoptions of their followers on complex network. Described in Watts threshold model, once the fraction of adopted neighbors of an individual exceeds his/her adoption threshold, he/she adopts the behavior. The mechanism of opinion leaders is introduced that an individual adopts the behavior when one opinion leader of this individual adopts the behavior. Two scenarios are put forward to choose the opinion leaders for each individual. One is that the individuals are randomly chosen from all individuals in the network. In this case, the mean degree of these chosen opinion leaders approximates the mean degree of network. The second is that the opinion leaders are chosen from the top $h\%$ of individuals with the highest degrees. The smaller value of $h\%$, the higher the mean degree of the opinion leaders. We systematically study the impacts of opinion leaders on behavior spreading for each scenario.

When the opinion leaders are randomly chosen from all individuals in the network, we find that the impacts of opinion leaders not only reduce the lowest mean degree of the network required for the global behavior adoption, also enlarge the highest mean degree of the network that the global behavior adoption can occur. Besides, the introduction of the opinion leaders accelerates the spread of behaviors, but it does not change the adoption order of the individuals. The theoretical predictions match with the simulation results.

When the opinion leaders are chosen from the top $h\%$ of individuals with the highest degree, we find there exists an optimal $h\%$ for the network with the lowest mean degree that the global behavior adoption can occur. The global behavior adoption becomes difficult when increasing the mean degree of opinion leaders. The impacts of opinion leaders in accelerating the spread of behaviors become less effective even can be ignored when the opinion leaders are with higher mean degree.

In this paper, we have studied how the opinion leaders--an important role in social contagion--of individuals impact the spread of behaviors. We purpose a mathematic model to describe the impacts of opinion leaders with the framework of Watts threshold model when the opinion leaders are random chosen from the network. Our results firstly present the influences of individual's opinion leaders on his/her behavior adoption in complex networks. The theory allows us to understand how the characteristic of opinion leaders shapes the behavior spreading. However, a number of questions still remain. For example, when the opinion leaders are chosen from the individuals with the high k-core~\cite{LiuY:2015}, high PageRank~\cite{Brin:1998} or high LeaderRank~\cite{Lv:2011}, will the current conclusion keep the same? Meanwhile, a more accurate and general theory method is still to be developed for the case of opinion leaders with high centrality. Our work may provide some suggestions in controlling or promoting the behavior adoption in real world, and also can stimulate more researches on social behavior spreading that takes into consideration both realistic spreading mechanisms and network topologies.\\

\section*{Acknowledgements}
This work was supported by the National Natural Science Foundation of
China under Grants Nos.~11575041, 61673086, the program of China Scholarships Council (Nos. 201606070059), the Natural Science Foundation of Shanghai (Grant No. 18ZR1412200) and the Fundamental Research Funds for the Central Universities (Grant
No.~ZYGX2015J153).

\appendix
\renewcommand\thesection{}
\section{ Derivation of $Eq. (5)$}

At the critical point, $\rho_{t-1}{\rightarrow}0$, thus we can
approximate ${(\rho_{t-1})}^m=0$, when $m>1$. Further,
Eq.~(1) can be written as
\begin{align}
\nonumber
L_t&=\sum_{k_0}Q(k_0)[1-{(1-\rho_{t-1})}^{k_0}]\\
\nonumber
&\approx\sum_{k_0}Q(k_0)\bigg[1-(1-k_0\rho_{t-1})\bigg]\\
&=\sum_{k_0}Q(k_0)k_0\rho_{t-1}. \tag{S1}
\end{align}

{\noindent}And Eq.~(4) can be written as
\begin{align}
\nonumber
q_t&=\rho_0+(1-\rho_0)\bigg[L_{t}+(1-L_{t})\sum_k{\frac{k}{z}p(k)}\\
\nonumber
&\times\sum_{m=0}^{k-1}B_{k-1,m}(q_{t-1})F\bigg(\frac{m}{k}\bigg)\bigg]\\
\nonumber
&\approx\rho_0+(1-\rho_0)\bigg[L_{t}+(1-L_{t})\sum_k{\frac{k}{z}p(k)}\\
&\times\bigg(F\bigg(\frac{0}{k}\bigg)+(k-1)q_{t-1}F\bigg(\frac{1}{k}\bigg)\bigg)\bigg]. \tag{S2}
\end{align}

{\noindent}Inserting Eq.~(S1) into Eq.~(S2) and using ${q_{t-1}}$ to approximate $\rho_{t-1}$ at the critical point, we have 
\begin{align}
\nonumber
q_t&\approx\rho_0+(1-\rho_0)\bigg[\sum_{k_0}Q(k_0)k_0\rho_{t-1}\\
\nonumber
&+\bigg(1-\sum_{k_0}Q(k_0)k_0\rho_{t-1}\bigg)\sum_k{\frac{k(k-1)}{z}p(k)}q_{t-1}F\bigg(\frac{1}{k}\bigg)\bigg]\\
&\approx\rho_0+(1-\rho_0)q_{t-1}\bigg[\sum_{k_0}Q(k_0)k_0+\sum_k{\frac{k(k-1)}{z}p(k)}F\bigg(\frac{1}{k}\bigg)\bigg].
\tag{S3}
\end{align}


\begin{references}

\bibitem{Barrat:2008}
A. Barrat, M. Barthelemy, and A. Vespignani, \emph{Dynamical processes on complex networks} (Cambridge university press, Cambridge, UK, 2008).

\bibitem{Newman:2010}
M. E. J. Newman, \emph{Networks: an introduction} (Oxford university press, Oxford, UK, 2010).

\bibitem{Castellano:2009}
C. Castellano, S. Fortunato, and V. Loreto, Rev. Mod. Phys. \textbf{81}, 591 (2009).

\bibitem{Pastor-Satorras:2015}
R. Pastor-Satorras, C. Castellano, P. Van Mieghem, and A. Vespignani, Rev. Mod. Phys.  \textbf{87}, 925 (2015).

\bibitem{WangWei:2017}
W. Wang, M. Tang, H. E. Stanley, and L. A. Braunstein, Rep. Prog. Phys. \textbf{80}, 036603 (2017).

\bibitem{Granovetter:1973}
M. Granovetter, Am. J. Sociol. \textbf{78}, 1360 (1973).

\bibitem{Centola:2007}
D. Centola, and M. Macy, Am. J. Sociol. \textbf{113} 702 (2007).

\bibitem{Watts:2002}
D. J. Watts, Proc. Natl. Acad. Sci. USA \textbf{99}, 5766 (2002).

\bibitem{Hackett:2011}
A. Hackett, S. Melnik, and J. P. Gleeson, Phys. Rev. E \textbf{83}, 056107 (2011).

\bibitem{Whitney:2010}
D. E. Whitney,  Phys. Rev. E \textbf{82}, 066110 (2010).

\bibitem{Gleeson:2008}
J. P. Gleeson, Phys. Rev. E \textbf{77}, 046117 (2008).

\bibitem{Nematzadeh:2014}
A. Nematzadeh, E. Ferrara, A. Flammini, and Y.-Y. Ahn, Phys. Rev. Lett. \textbf{113}, 088701 (2014).

\bibitem{Hurd:2013}
T. R. Hurd, and J. P. Gleeson, J. Complex Networks \textbf{1}, 25 (2013).

\bibitem{Yagan:2012}
O. Yagan and V. Gligor, Phys. Rev. E \textbf{86}, 036103 (2012).

\bibitem{Brummitt:2012}
C. D. Brummitt, K.-M. Lee, and K.-I. Goh, Phys. Rev. E \textbf{85},
045102(R) (2012).

\bibitem{Lee:2014}
K.-M. Lee, C. D. Brummitt, and K.-I. Goh, Phys. Rev. E \textbf{90},
062816 (2014).

\bibitem{Takaguchi:2013}
T. Takaguchi, N. Masuda, and P. Holme, PLoS ONE \textbf{8}, e68629 (2013).

\bibitem{Glesson:2007}
J. P. Glesson and D. J. Cahalane, Phys. Rev. E \textbf{75}, 056103 (2007).

\bibitem{Lee:2017}
E. Lee and P. Holme, Phys. Rev. E \textbf{96}, 012315 (2017).

\bibitem{Kobayashi:2015}
T. Kobayashi, Phys. Rev. E \textbf{92}, 062823 (2015).

\bibitem{Ruan:2015}
Z. Ruan, G. Iniguez, M. Karsai, and J. Kert\'{e}sz,  Phys. Rev. Lett. \textbf{115}, 218702 (2015).

\bibitem{Huang:2016}
W. M. Huang, L. J. Zhang, X. J. Xu, and X. Fu, Sci. Rep. \textbf{6}, 23766 (2016).

\bibitem{Lu:2011}
L. L\"u, D. B. Chen, and T. Zhou, New J. Phys. \textbf{13}, 123005 (2011).

\bibitem{Zheng:2013}
M. Zheng, L. L\"{u}, and M. Zhaos, Phys. Rev. E \textbf{88}, 012818 (2013).

\bibitem{Liu:2016}
Q. H. Liu, W. Wang, M. Tang, and H. F. Zhang, Sci. Rep. \textbf{6}, 25617 (2016).

\bibitem{Liu:2017}
Q. H. Liu, W. Wang, M. Tang, T. Zhou, and Y. C. Lai, Phys. Rev. E \textbf{95}, 042320 (2017).

\bibitem{Wang:2015}
W. Wang, M. Tang, H. F. Zhang, and Y. C. Lai, Phys. Rev. E \textbf{92}, 012820 (2015).

\bibitem{Zhang:2013}
J. Zhang, B. Liu, J. Tang, T. Chen, and J. Li, \emph{Proceedings of the Twenty-Three International Joint Conference on Artificial Intelligence} Beijing, China. Menlo Park, California, USA: AAAI Press (2013).

\bibitem{Hodas:2014}
N. O. Hodas and K. Lerman, Sci. Rep. \textbf{4}, 4343 (2014).

\bibitem{Cozzo:2013}
E. Cozzo, R. A. Banos, S. Meloni, and Y. Moreno, Phys. Rev. E \textbf{88}, 050801 (2013).

\bibitem{Liu:2018-1}
Q. H. Liu, L. F. Zhong, W. Wang, T. Zhou and H. E. Stanley, Chaos \textbf{28}, 013120 (2018).

\bibitem{Liu:2018-2}
Q. H. Liu, W. Wang, S. M. Cai, M. Tang and Y. C. Lai, Phys. Rev. E \textbf{97}, 022311 (2018).

\bibitem{Walter:2014}
W. Quattrociocchi, G. Caldarelli, and A. Scala, Sci. Rep. \textbf{4}, 4938 (2014).

\bibitem{Rogers:1962}
E. M. Rogers and D. G. Cartano, Public Opin. Q. \textbf{26}, 435 (1962).

\bibitem{Myers:1972}
J. H. Myers and T. S. Robertson, J. Mark. Res. \textbf{9}, 41 (1972).

\bibitem{Richins:1988}
M. L. Richins and T. Root-Shaffer, ACR North American Advances \textbf{15}, 32 (1988).

\bibitem{Iyengar:2011}
R. Iyengar, C. Van den Bulte, and T. W. Valente, Market. Sci. \textbf{30}, 195 (2011).

\bibitem{Lu:2016}
L. L\"u, D. Chen, X. L. Ren, Q. M. Zhang, Y. C. Zhang and T. Zhou, Phys. Rep. \textbf{650}, 1 (2016).

\bibitem{kelly:1991}
J. A. Kelly, J. S. St Janet, Y. E. Lawrence. L. Diaz, S. Yvonne, \emph{et al}, Am. J. Public Health \textbf{81}, 168 (1991).

\bibitem{lomas:1991}
J. Lomas, E. Murray, M. A. Geoffrey, J. H. Walter, V. Eugene, \emph{et al},
JAMA \textbf{265}, 2202 (1991).

\bibitem{Earp:2002}
J. A. Earp, E. Eng, M. S. O'Malley, M. Altpeter, G. Rauscher \emph{et al}, Am. J. Public Health \textbf{92}, 646 (2002)

\bibitem{Park:2013}
C. S. Park, Comput. Human Behav. \textbf{29}, 1641 (2013).

\bibitem{Hooper:2010}
P. L. Hooper, H. S. Kaplan, and J. L. Boone, J. Theor. Biol. \textbf{265}, 633 (2010).

\bibitem{Gavrilets:2016}
S. Gavrilets, J. Auerbach and M. Van Vugt, Sci. Rep. \textbf{6}, 29704 (2016).

\bibitem{Rogers:2010}
E. M. Rogers, \emph{Diffusion of innovations} (Simon and Schuster, 2010).

\bibitem{Cho:2012}
Y. Cho, J. Hwang, and D. Lee, Technol. Forecase Soc. Change \textbf{79}, 97 (2012).

\bibitem{Eck:2011}
P. S. Van Eck, W. Jager and P. S. Leeflang, J. Prod. Innov. Manag. \textbf{28}, 187 (2011).

\bibitem{Catanzaro:2005}
M. Catanzaro, M. Bogu\~{n}\'{a} and R. Pastor-Satorras, Phys. Rev. E \textbf{71}, 027103 (2005).

\bibitem{Valente:2007}
T. W. Valente and P. Pumpuang, Health Educ. Behav. \textbf{34}, 881 (2007).

\bibitem{ERDdS:1959}
P. Erd\"{o}s and A. R\'{e}nyi, Publ. Math. \textbf{6}, 290 (1959).

\bibitem{LiuY:2015}
Y. Liu, M. Tang, T. Zhou and Y. Do, Sci. Rep. \textbf{5}, 9602 (2015).


\bibitem{Brin:1998}
S. Brin, L. Page, Comput. Netw. ISDN Syst. \textbf{30}, 107 (1998).

\bibitem{Lv:2011}
L. L\"{u}, Y. C. Zhang, C. H. Yeung and T. Zhou, PloS one \textbf{6}, e21202 (2011).



\end{references}
\end{document}